\documentclass[showpacs,preprintnumbers,aps]{revtex4}
\begin{document}
\preprint{USM-KR2}
\newcommand{\ba}{\begin{eqnarray}}
\newcommand{\ea}{\end{eqnarray}}
\newcommand{\be}{\begin{equation}}
\newcommand{\ee}{\end{equation}}
\newcommand{\bib}{\bibitem}
\newcommand{\ed}{\end{document}}
\newcommand{\nn}{\nonumber\\}
\newcommand{\fr}{\frac}
\newcommand{\wt}{\widetilde}

\title{On condensation of topological defects and confinement}
\author{Patricio Gaete}
\email {patricio.gaete@usm.cl} \affiliation{Departamento de
F\'{\i}sica, Universidad T\'ecnica F. Santa Mar\'{\i}a,
Valpara\'{\i}so, Chile}
\author{Clovis Wotzasek}
\email {clovis@if.ufrj.br}
\address{Instituto de F\'\i sica, Universidade Federal do Rio de
Janeiro, 21945, Rio de Janeiro, Brazil}
\date{\today}
\begin{abstract}
We study the static quantum potential for a theory of anti-symmetric
tensor fields that results from the condensation of topological
defects, within the framework of the gauge-invariant but
path-dependent variables formalism. Our calculations show that the
interaction energy is the sum of a Yukawa and a linear potentials,
leading to the confinement of static probe charges.
\end{abstract}
\thanks{One of us (PG) would like to dedicate this work to the memory of Jian-Jun Yang}
\maketitle

\section{INTRODUCTION}

One of the fundamental and long-standing issues of theoretical
physics whose solution has evaded complete comprehension despite
an intense effort for already many decades is that of the
confinement for the fundamental constituents of matter. On the
other hand a great many deal of progress has been made towards
making an unequivocal distinction between the apparently related
phenomena of screening and confinement. In fact such distinction
is of considerable importance in our present understanding of
gauge theories. Field theories that yield the linear potential are
very important to particle physics, since those theories may be
used to describe the confinement of quarks and gluons and be
considered as effective theories of quantum chromodynamics. Due to
intense interest mainly in string related topics, these studies
have been extended out of the four dimensional domain and extended to
theories of antisymmetric tensors of arbitrary ranks in arbitrary
space-time dimensions that appear as low-energy effective field
theories of strings. This has also helped us to gain insights over
the mechanisms of confinement in different contexts.

It is the main purpose of this work to study the confinement versus
screening properties of some theories of massless antisymmetric tensors,
magnetically and electrically coupled to topological defects that
eventually condense, as a consequence of the Julia--Toulouse mechanism \cite{JT}.
This mechanism is the dual to the Higgs mechanism and has been shown by
Quevedo and Trugenberger \cite{QT} to lead to a concrete massive antisymmetric theory with a jump of rank.
We clearly show that in the presence of two tensor fields, the condensation induces, not only a mass term
and a jump of rank for one of the tensors but a BF coupling
will also become manifest which will be responsible for the change from the screening
to the confining phase of the theory.

In another circumstances antisymmetric tensor theories have
been studied in the past because they are the natural extension of
abelian gauge theories, our basic model for a gauge theory, with
which they share very interesting properties, particularly the
essential property of duality, leading to the strong--weak coupling
mapping, known to exist in the electromagnetism. Such features are
known to exist also in antisymmetric tensor theories in any dimension
which will be essential for us in the sequel. Antisymmetric
tensors also appear very naturally in supersymmetric field
theories and in string theories where they play an important role
in the realization of the various strong-weak coupling dualities
among string theories.

An antisymmetric tensor of rank $p$ couples naturally to an
elementary extended $p-1$ dimensional object, a $(p-1)$-brane
since its world-hypersurface is a $p$-dimensional object. However,
if the antisymmetric tensors are compact fields, there may also
appear defects or solitonic excitations in the underlying theory.
Defects are classical solutions of the equations of motion and may
be classified as topological or nontopological. Topological
defects appear in models that support spontaneous symmetry
breaking. They are important also in Cosmology and Condensed
Matter Physics. Condensation of topological defects may drive
phase transitions, particularly from the screening or Coulomb
phases to the confining phase, which is our main interest in this
paper. The prototype of this phenomenon is the well known
Kosterlitz--Thouless transition in two space dimensions driven by
vortex condensation. An important question in the analysis of
phase transitions induced by topological defects regards the
conditions for a topological defect to condense and for which
values of parameters like temperature and coupling constants. Such
questions have been tackled mostly by lattice techniques.

Another important issue regards the nature of the new phase with a
finite condensate of topological defects. It is this last aspect,
in $D=d+1$ space-time dimensions for generic antisymmetric tensor
field theories, that is of importance for us in this paper. This
issue was discussed long time ago by Julia and Toulouse \cite{JT}
in the framework of ordered solid-state media and more recently in
the relativistic context by Quevedo and Trugenberger \cite{QT}. The
basic idea in Ref.\cite{JT} was to consider models with
non-trivial homotopy group able to support stable topological
defects characterized by a length scale $r$ \footnote{ This scale
may be written in terms of a mass parameter $M$ to be considered
as a cut-off for the low-energy effective field theory as
$r=1/M$.}. The long wavelength fluctuations of the continuous
distribution of topological defects are the new hydrodynamical
modes for the low-energy effective theory that appear when
topological defects condense. In \cite{JT} there is a clear cut
algorithm to identify these new modes in the framework of ordered
solid-state media. However, due to the presence of non-linear
terms in the topological currents, the lack of relativistic
invariance and the need to introduce dissipation terms it becomes
difficult to write down an action for the phase with a condensate
of topological defects in this framework.

In the relativistic context none of the above problems is present.
This allowed Quevedo and Trugenberger to show that the
Julia--Toulouse prescription can be made much more precise,
leading to an explicit form for the action in the finite
condensate phase, for generic compact antisymmetric field
theories. In this context the Julia--Toulouse mechanism is the
natural generalization of the confinement phase for a vector gauge
field.

In this paper we make use of the Julia--Toulouse mechanism, as
presented by Quevedo and Trugenberger, to study the low-energy field
theory of a pair of massless anti-symmetric tensor fields, say
$A_p$ and $B_q$ with $p+q+2=D$, coupled electrically and
magnetically to a large set of $(q-1)$-branes, characterized by
charge $e$ and a Chern-Kernel $\Lambda_{p+1}$ \cite{HL}, that eventually
condense. It is shown that the effective theory that results
displays the confinement property by computing explicitly the
effective potential for a pair of static, very massive point
probes.

Basically, we are interested in studying the Julia--Toulouse
mechanism in model field theories involving $B_q$ and $A_p$,
electrically and magnetically coupled to a $(q-1)$-brane,
respectively, according to the following action \ba \label{R10}
{\cal S} &=&  \int \frac 12 \frac{(-1)^q}{(q+1)!}
\left[H_{q+1}(B_q) \right]^2 + e\, B_q J^{q}(\Lambda) +  \frac 12
\frac{(-1)^p}{(p+1)!}  \left[F_{p+1}(A_p) - e
\Lambda_{p+1}\right]^2 \ea and consider the condensation
phenomenon when the Chern-Kernel $\Lambda_{p+1}$ becomes the new
massive mode of the effective theory. Our compact notation here
goes as follows. The field strength reads
$F_{p+1}\left(A_p\right)= F_{\mu_1 \mu_2 \ldots
\mu_{p+1}}=\partial_{[\mu_1}A_{\mu_2\cdots\mu_{p+1}]}$ and
$\Lambda_{p+1}=\Lambda_{\mu_1\cdots\mu_{p+1}}$ is a totally
anti-symmetric object of rank ($p+1$).
The conserved current $J^q(\Lambda)$ is given by a
delta-function over the world-volume of the ($q-1$)-brane
\cite{Kleinert:kx}. This conserved current may be rewritten in
terms of the kernel $\Lambda_{p+1}$ as \be J^q(\Lambda) = \frac
1{(p+1)!} \epsilon^{q,\alpha ,p+1}\partial_\alpha \Lambda_{p+1}\;
, \ee and $\epsilon^{q,\alpha ,p+1} =
\epsilon^{\mu_1\ldots\mu_q,\alpha ,\nu_1\ldots\nu_{p+1}}$. This
notation will be used in the discussion of the Julia-Toulouse
mechanism in the next section as long as no chance of confusion
occurs. We will show in the next section how the
Quevedo-Trugenberger prescription is used to constructed the
effective interacting action,  in the condensed phase, between the
anti-symmetric tensor field $B_q$ and the degrees of freedom of
the condensate $\Lambda_{p+1}$. In Section III we study the
confinement properties of this effective action, after the
condensate is integrate out, by computing the effective quantum potential
for a pair of static probe charges within the framework of the gauge-invariant but
path-dependent variables formalism. In particular, we shall be interested in the dependence of the confinement properties with the condensation parameters coming from the Julia--Toulouse mechanism.

\section{The Effective Action}

\subsection{The Julia--Toulouse Mechanism}

We begin with a discussion of the Julia--Toulouse mechanism
\cite{JT} in the relativistic context as originally developed in
Ref.\cite{QT} which we follow closely. Consider a generic
field theory in a $D$-dimensional space-time whose symmetry group
$G$ is spontaneously broken down to $H$. Topological defects may
arise in this theory according to the values taken by the homotopy
group of the quotient-space $\Pi_h\left(G/H\right)$. For $h<d$,
a non-trivial $\Pi_h\left(G/H\right)$ will lead to
$(d-h-1)$--dimensional solitons  while instantons appears in the
Euclidean version of the model when $h=d$. The characteristic
sizes of these extended classical solutions with finite-energy are
$r_i=1/M_i$, where $M_i$ are mass parameters associated with the
spontaneous symmetry breaking process. For
ordered solid-state media \cite{JT} the low-energy excitations are
generically described by field theories for some order parameter
as in the Ginzburg--Landau theory. For the case at hand, the
effective low-energy theory that has symmetry group $H$, is
meaningful only on scales much bigger than $max\{r_i \}$ and, most
important, experiences topological defects essentially as
$d-h-1$--dimensional singularities in the space $R^d$ (for
solitons) or point singularities in $R^{d+1}$ (for instantons). To
our interest here the important fields are the anti-symmetric
tensors of rank $(d-h)$ and $(h-1)$ that are able to couple
electrically and magnetically, respectively, to the singularities.
Besides, the most important point to stress here is that the
effective action for this low-energy theory is then well-defined
only outside these singularities.

As discussed in the introduction we wish to focus on
anti-symmetric tensor field theories which are simple, yet highly
non-trivial generalization of the usual Maxwell theory.
Furthermore, we consider compact antisymmetric field theories,
which are the generalizations of the compact QED, as put forward
in \cite{Polyakov,PO}, to higher-rank tensor theories.
In compact antisymmetric field theories $p$-branes
appear as topological defects of the original theory and their
world-volume can be viewed as closed $(d - h)$-dimensional
singularities excluded from the (space-time) domain of the model.
They constitute the charges of the effective theory with strength
$e_i$ and Chern-Kernels $\Lambda_h^{(i)}$ such that
$J^{d-h}_i=\epsilon^{d-h,1,h}\,\partial \Lambda_h^{(i)}$. These
charges together with the anti-symmetric tensors are the basic
building blocks in the construction of the low-energy effective
field theory outside these singularities.

To understand the mechanism proposed by Julia and Toulouse one
has to observe that the presence of these singularities induces
non-trivial homology cycles for the antisymmetric tensor fields.
There are then topological quantum number given in terms of the
fields and the singularities, that are essentially
$(d-h-1)$--dimensional holes defined on $D$--dimensional Minkowski
space-time while instantons would correspond to a model on
$(d+1)$--dimensional Euclidean space with holes at the location of
the instantons. To realize the significance of this we consider
the intersection of these singularities with an $(h +
1)$-dimensional hyperplane $\Sigma_{h+1}$ perpendicular to it.
Consider a sphere $S_h$ on $\Sigma_{h+1}$ such that it isolates
one of the two intersection points with the singularities
world-volume (this is a $(d - h)$--dimensional object). To see the
topological quantum numbers, given in terms of $S_h$, generated by
stable topological defects, for which the relevant homotopy groups
are $\Pi_h \left(G/H\right)=Z$, we proceed as follows. Let us
define an $h$-form $\Omega_h$ which is exact outside $S_h$ whose
components read (except for normalization)
$\Omega_{\mu_1\cdots\mu_h} =
\partial_{[\mu_1}\phi_{\mu_2\cdots\mu_{h}]}$ where $\phi_{h-1}$ is an
$(h-1)$-anti-symmetric tensor field. Then there exists topological
invariants that can be written as $\Phi =
{\int_{S_h}\Omega_h}d\sigma^h$ where the compact notation of the
preceding section has been invoked.

Let us consider next the effect of many topological defects on the
low-energy effective theory since topological defects can condense
leading to drastic changes in the infra-red structure of the
underlying theory \cite{PO}. Therefore, the relevant question
to address regards the change of the physical scenario when the
number of topological defects grows making the manifold on which
the low-energy effective action is defined to become very
complicated. It may happens then that, for a certain range of
parameters, dynamics favors the formation of a macroscopic number
of topological excitations (the condensate) with a finite density.
The condensate is then described by new long-lived modes. To
identify the additional hydrodynamical modes of a solid state
medium due to the continuous distribution of topological defects
one uses the Julia--Toulouse theory, while in the framework of
relativistic field theories, Quevedo and Trugenberger have provided
a prescription to show that these modes are manifest as additional
fields in the new phase of the low-energy theory at finite density
of topological defects.

The new massive low-energy modes in the phase with a finite
density of topological defects are obtained by promoting the
closed form $\Omega_h$ defined outside the complicated manifold
that includes the plethora of defects to a fundamental form
defined everywhere on $R^{d+1}$. After condensation the original
potential $\phi_{h-1}$ becomes meaningless. In particular, for the
model discussed in Section I, this role is played by the
Chern-Kernel $\Lambda_h$, now promoted to the category of
propagating tensor, that becomes the smooth field on $R^{d+1}$
describing the conserved fluctuations of the continuous
distribution of topological defects which constitute the new
hydrodynamical modes.

Although the JT algorithm becomes problematic in the framework of
ordered solid state media due to non-linearities and the need of
introducing dissipation terms, Quevedo and Trugenberger have shown
that it leads to simple demands in the study of compact
antisymmetric tensor theories, where it also produces naturally
the effective action for the new phase. They observe that when the
$(d-h-1)$-branes condense this generates a new scale $\Delta$ that
is related to the average density $\rho $ of intersection points
of the $(d-h)$-dimensional world-hypersurfaces of the condensed
branes with any $(h+1)$-dimensional hyperplane. The four
requirements made by Quevedo and Trugenberger over the action
candidate to describe effectively the dense phase are: (i) an
action built up to two derivatives in the new field possessing
(ii) gauge invariance, (iii) relativistic invariance  and, most
important, (iv) the need to recover the original model in the
limit $\Delta \to 0$.  One is therefore led to consider the action
for the condensate as \ba \label{RM15} {\cal S}_{\Omega}=\int
\frac {(-1)^h }{2 \Delta^2(h+1)!} \left[{F_{h+1}(\Omega_h)}
\right]^2 -\frac {(-1)^h\, h!}{2\, e^2} \left[\Omega_h - H_h
(\phi_{h-1})\right]^2 \ea where $H_{\mu_1\cdots\mu_h} =
\partial_{[\mu_1}\phi_{\mu_2\cdots\mu_{h}]}$ and
the underlying gauge invariance is manifest by the simultaneous
transformations $\Omega_{\mu_1\cdots\mu_h} \to
\Omega_{\mu_1\cdots\mu_h} +
\partial_{[\mu_1}\psi_{\mu_2\cdots\mu_h]}$ and
$\phi_{\mu_1\cdots\mu_{h-1}} \to \phi _{\mu_1\cdots\mu_{h-1}}+
\psi_{\mu_1\cdots\mu_{h-1}}$. Upon fixing this gauge invariance
one can drop all considerations over $\phi_{h-1}$ after absorbing
$H_h(\phi_{h-1})$ into $\Omega_h$, so that the action describes
the exact number of degrees of freedom of a massive field whose
mass parameter reads $m=\Delta/e$. This process, named as
Julia-Toulouse mechanism by Quevedo and Trugenberger, is the process
dual to the well known Higgs mechanism of electric charges. Here
on the other hand, the new modes generated by the condensation of
magnetically coupled topological defects absorbs the original
variables of the effective field theory, thereby acquiring a mass
while in the Higgs mechanism it is the original field that
incorporates the degrees of freedom of the electric condensate to
acquire mass. This difference explains the change of rank in the
JT mechanism that is not present in the Higgs process.

In the limit $\Delta \to 0$ the only relevant field configurations
are those that satisfy the constraint $F_{h+1}(\Omega_h) =0$ whose
solution reads $\Omega_{\mu_1\cdots\mu_h} =
\partial_{[\mu_1}\psi_{\mu_2\cdots\mu_{h}]}$ where $\psi_{h-1}$ is an
$(h-1)$-anti-symmetric tensor field. The field $\psi_{h-1} $ can
then be absorbed into $\phi_{h-1} $ this way recovering the
original low-energy effective action before condensation.

\subsection{The Action in the Condensed Phase}

We are now ready to discuss the consequences of the JT mechanism
over the action (\ref{R10}).  The distinctive feature is that
after condensation the Chern-Kernel $\Lambda_{p+1}$ is elevated to
the condition of propagating field. The new degree of freedom
absorbs the degrees of freedom of the tensor $A_p$ this way
completing its longitudinal sector. The new mode is therefore
explicitly massive. Since $A_p\to \Lambda_{p+1}$ there is a change
of rank with dramatic consequences. The last term in (\ref{R10}),
displaying the magnetic coupling between the field-tensor
$F_{p+1}(A_p)$ and the $(q-1)$-brane, becomes the mass term for
the new effective theory in terms of the tensor field
$\Lambda_{p+1}$ and a new dynamical term is induced by the
condensation.
Another important feature is that the minimal coupling
of the $B_q$ tensor becomes responsible for another contribution
for the mass, this time of topological nature. Indeed the second
term (\ref{R10}) becomes an interacting ``$BF$--term" in the form
``$B\wedge F(\Lambda)$" term between the remaining propagating
modes, inducing the appearance of topological mass, in addition to
the induced condensate mass.  The final result reads \ba
\label{RM10} {\cal S}_{cond} = \int  \frac{(-1)^q}{2(q+1)!}
\left[H_{q+1}(B_q) \right]^2 + e\, B_q
\epsilon^{q,\alpha,p+1}\partial_\alpha \Lambda_{p+1} + \frac
{(-1)^{p+1}}{2 (p+2)!} \left[F_{p+2}(\Lambda_{p+1}) \right]^2
-\frac {(-1)^{p+1}\, (p+1)!}2 m^2 \Lambda_{p+1}^2 \ea where $m =
\Delta/ e$.

Recall that the initial theory, before condensation, displayed two
independent fields coupled to a $(q-1)$--brane. The nature of the
two couplings were however different with important consequences.
The $A_p$ tensor, that was magnetically coupled to the brane, was
then absorbed by the condensate after phase transition. On the
other hand, the electric coupling, displayed by the $B_q$ tensor,
became a ``$B\wedge F(\Lambda)$" topological term after
condensation.

There has been a drastic change in the physical scenario. To show
that the new systems displays a confining phase is the goal of
this work.  In this section we want to obtain an effective action
for the $B_q$ tensor. To this end we shall next integrate out the
field describing the condensate. The implications of the resulting
effective action will be studied in the next section.

To integrate out the condensate field $\Lambda$ we rewrite this
sector of the action as, \ba \label{RM100} {\cal S}_{\Lambda} &=&
\int   \frac {(-1)^{p+1}}{2 (p+2)!} \left[F_{p+2}(\Lambda_{p+1})
\right]^2 -\frac {(-1)^{p+1}\, (p+1)!}2 m^2 \Lambda_{p+1}^2 +  e\,
B_q \epsilon^{q,\alpha,p+1}\partial_\alpha \Lambda_{p+1}\nn &=&
\int \frac{(-1)^{p+2}(p+1)!}{2}\, \Lambda_{p+1}\left(\Delta^2
+m^2\right)\Lambda^{p+1} +  e\, B_q
\epsilon^{q,\alpha,p+1}\partial_\alpha \Lambda_{p+1} \ea where we
have made use of (i) the identity
\begin{eqnarray}
\label{identity}
\epsilon_{\mu_1\cdots\mu_p\nu_1\cdots\nu_q}\,\epsilon^{\mu_1\cdots\mu_p\nu_1\cdots\nu_q}
= (-1)^{D+1}\, p! \,
\delta^{\tilde\nu_1}_{[\nu_1}\cdots\delta^{\tilde\nu_q}_{\nu_q]}
\end{eqnarray}
which in our compact notation reads
\begin{eqnarray}
\label{compactidentity}
    \epsilon_{p,q} \,\epsilon^{p,q}=
    (-1)^{D+1}\, p! \, \delta^{\;\tilde q}_{[q]}
\end{eqnarray}
and (ii) of an integration by parts, such that
\begin{eqnarray}
\left(\epsilon^{q,\alpha,p+1}\partial_\alpha\Lambda_{p+1}\right)^2
= (-1)^D\, q!\, p!\, \Lambda_{p+1} \,(\Delta^2) \,\Lambda^{p+1}
\end{eqnarray}
Next we solve the equations of motion
\begin{eqnarray}
    \frac{\delta{\cal S}_{\Lambda}}{\delta\Lambda_{p+1}} = 0
\end{eqnarray}
to obtain
\begin{eqnarray}
\label{solution} \Lambda_{p+1} = \frac{e \,(-1)^{p+1}\,
\wp(\epsilon\partial)}{(p+1)!} \frac{1}{\Delta^2 +
m^2}\;\epsilon_{p+1,\alpha,q}\,\partial^\alpha \, B^q
\end{eqnarray}
where $\wp(\epsilon\partial)= (-1)^{p(q+1)}$ is the parity of the
generalized curl operator, in the sense that
\begin{eqnarray}
\int \psi_p\,\epsilon^{p,\alpha,q}\partial_\alpha \, \phi_q
=\wp(\epsilon\partial) \int \phi_q
\,\epsilon^{q,\alpha,p}\partial_\alpha \,\psi_p\, .
\end{eqnarray}
Substituting (\ref{solution}) back into the action (\ref{RM100})
and using that,
\begin{eqnarray}
\left(\epsilon_{p+1,\alpha,q}\partial^\alpha B^q\right)^2 =
(-1)^{D+1}
 \frac{(p+1)!}{(q+1)1} H_{q+1}(B_q) H^{q+1}(B_q)
\end{eqnarray}
gives, after the inclusion of the free term for the $B_q$ tensor,
our final effective theory as
\begin{eqnarray}
\label{acaoeffetiva}
{\cal S}_{eff} =\int \frac{(-1)^{q+1}}{2\, (q+1)!} H_{q+1}(B_q)
\left(1 +\frac{e^2}{\Delta^2 + m^2} \right)H^{q+1}(B_q)
\end{eqnarray}

In the next section we shall examine the screening versus
confinement issue. To this end we shall consider a specific
example involving two Maxwell tensors coupled electrically
and magnetically to a point-charge (a zero-brane) such that after
the condensation we end up with a Maxwell and a Kalb-Ramond field
(the condensate) coupled topologically to each other, besides the
presence of an explicit mass term for the Kalb-Ramond.

\section{Interaction Energy}

Our aim in this Section is to calculate the interaction
energy for the effective theory computed above between external probe sources in an specific model. To do this, we will compute the expectation value of the energy operator
$H$ in the physical state $\left| \Phi  \right\rangle$ describing the sources, which we
will denote by $ \left\langle H \right\rangle _\Phi$. It is worth mentioning that our starting
point, Eq.(\ref{RM10}), with $p=q=1$ is the Lagrangian obtained in \cite{Deguchi}:
\begin{equation}
{\cal L} = \frac{1}{{12}}H_{\mu \nu \rho } H^{\mu \nu \rho }  -
\frac{1}{4}m^2 \Lambda_{\mu \nu }\Lambda^{\mu \nu } - \frac 14 F_{\mu \nu } F^{\mu \nu }  
- \frac 12\, e \Lambda_{\mu \nu }F^{\mu \nu }  - A_0 J^0, \label{KR10}
\end{equation}
where the Kalb-Ramond field $\Lambda_{\mu\nu}$ carries the degrees of freedom of the condensate, as discussed at the end of the last Section.
Here $H_{\mu \nu \rho }  = \partial _\mu  \Lambda_{\nu \rho }  +
\partial _\rho  \Lambda_{\mu \nu }  + \partial _\nu  \Lambda_{\rho \mu }$,
$F_{\mu \nu }  = \partial _\mu  A_\nu   - \partial _\nu  A_\mu$
and $J^0$ is an external current. As stated, our objective will be to
calculate the potential energy for this theory. As in the previous
subsection, the first step in this direction is to carry out the
integration over $\Lambda_{\mu \nu }$ in Eq.(\ref{KR10}). This allows us
to write the following effective Lagrangian
\begin{equation}
{\cal L} =  - \frac{1}{4}F_{\mu \nu } \left( {1 + \frac{{e^2
}}{{\triangle^2 + m^2 }}} \right)F^{\mu \nu }  - A_0 J^0.
\label{KR15}
\end{equation}
which is a particular case of (\ref{acaoeffetiva}). We observe, either from (\ref{KR10}) or from (\ref{KR15}), that the limits $e\to 0$ or $m\to 0$ are well defined and lead (from the point of view of the probe charges) to a pure Maxwell theory or to a (topologically) massive model. Since the probe charges only couple to the Maxwell fields, the Kalb-Ramond condensate will not contribute to their interaction energy in the first case because in the limit where the parameter $e\to 0$ the Maxwell field and the condensate decouple. The second limit means that we are back to the non-condensed phase. As so the confinement of the probe charges are expected to disappear being taken over by an screening phase controlled by the parameter $e$ playing the role of topological mass. 

Once this is done, the canonical quantization of this theory from
the Hamiltonian point of view follows straightforwardly. The canonical
momenta read $\Pi ^\mu   = - \left( {1 + \frac{{e^2 }}{{\Delta ^2
+ m^2 }}} \right)F^{0\mu }$ with the only nonvanishing canonical
Poisson brackets being
\begin{equation}
\left\{ {A_\mu  \left( {t,x} \right),\Pi ^\nu  \left( {t,y}
\right)} \right\} = \delta _\mu ^\nu  \delta \left( {x - y}
\right). \label{KR20}
\end{equation}
Since $\Pi_0$ vanishes we have the usual primary constraint
$\Pi_0=0$, and $\Pi ^i  = \left( {1 + \frac{{e^2 }}{{\Delta ^2 +
m^2 }}} \right)F^{i0}$. The canonical Hamiltonian is thus
\begin{equation}
H_C  = \int {d^3 } x\left\{ { - \frac{1}{2}\Pi ^i \left( {1 +
\frac{{e^2 }}{{\Delta ^2  + m^2 }}} \right)^{ - 1} \Pi _i  + \Pi
^i \partial _i A_0  + \frac{1}{4}F_{ij} \left( {1 + \frac{{e^2
}}{{\Delta ^2  + m^2 }}} \right)F^{ij}  + A_0 J^0 }
\right\}.\label{KR25}
\end{equation}
Time conservation of the primary constraint $ \Pi _0$ leads to the
secondary Gauss-law constraint
\begin{equation}
\Gamma _1 \left( x \right) \equiv \partial _i \Pi ^i - J^0 =
0.\label{KR30}
\end{equation}
The preservation of $\Gamma_1$ for all times does not give rise to
any further constraints. The theory is thus seen to possess only
two constraints, which are first class, therefore the theory
described by $(\ref{KR15})$ is a gauge-invariant one. The extended
Hamiltonian that generates translations in time then reads $H =
H_C  + \int {d^3 } x\left( {c_0 \left( x \right)\Pi _0 \left( x
\right) + c_1 \left( x \right)\Gamma _1 \left( x \right)}
\right)$, where $c_0 \left( x \right)$ and $c_1 \left( x \right)$
are the Lagrange multiplier fields. Moreover, it is
straightforward to see that $\dot{A}_0 \left( x \right)= \left[
{A_0 \left( x \right),H} \right] = c_0 \left( x \right)$, which is
an arbitrary function. Since $ \Pi^0 = 0$ always, neither $ A^0 $
nor $ \Pi^0 $ are of interest in describing the system and may be
discarded from the theory. Then, the Hamiltonian takes the form
\begin{equation}
H = \int {d^3 x\left\{ { - \frac{1}{2}\Pi _i \left( {1 +
\frac{{e^2 }}{{\Delta ^2  + m^2 }}} \right)^{ - 1} \Pi ^i  +
\frac{1}{4}F_{ij} \left( {1 + \frac{{e^2 }}{{\Delta ^2  + m^2 }}}
\right)F^{ij}  + c\left( x \right)\left( {\partial _i \Pi ^i  -
J^0 } \right)} \right\}}, \label{KR35}
\end{equation}
where $c(x) = c_1 (x) - A_0 (x)$.

The quantization of the theory requires the removal of nonphysical
variables, which is done by imposing a gauge condition such that
the full set of constraints becomes second class. A convenient
choice is found to be \cite{Pato}
\begin{equation}
\Gamma _2 \left( x \right) \equiv \int\limits_{C_{\xi x} } {dz^\nu
} A_\nu \left( z \right) \equiv \int\limits_0^1 {d\lambda x^i }
A_i \left( {\lambda x} \right) = 0, \label{KR40}
\end{equation}
where  $\lambda$ $(0\leq \lambda\leq1)$ is the parameter
describing the spacelike straight path $ x^i = \xi ^i  + \lambda
\left( {x - \xi } \right)^i $, and $ \xi $ is a fixed point
(reference point). There is no essential loss of generality if we
restrict our considerations to $ \xi ^i=0 $. In this case, the
only nonvanishing equal-time Dirac bracket is
\begin{equation}
\left\{ {A_i \left( x \right),\Pi ^j \left( y \right)} \right\}^ *
=\delta{ _i^j} \delta ^{\left( 3 \right)} \left( {x - y} \right) -
\partial _i^x \int\limits_0^1 {d\lambda x^j } \delta ^{\left( 3
\right)} \left( {\lambda x - y} \right). \label{KR45}
\end{equation}
In passing we recall that the transition to quantum theory is made
by the replacement of the Dirac brackets by the operator
commutation relations according to
\begin{equation}
\left\{ {A,B} \right\}^ *   \to \left( { - i} \right)\left[ {A,B}
\right]. \label{KR50}
\end{equation}

We now turn to the problem of obtaining the interaction energy
between pointlike sources in the model under consideration. The state $\left| \Phi  \right\rangle$
representing the sources is obtained by operating over the vacuum with creation/annihilation operators. We want to stress that, by construction, such states are gauge invariant. In the case at hand we consider the gauge-invariant stringy $\left|{\overline \Psi  \left( \bf y \right)\Psi \left( {\bf y^ \prime }
\right)} \right\rangle$, where
a fermion is localized at ${\bf y}\prime$ and an antifermion at $
{\bf y}$ as follows \cite{Dirac2}, 
\begin{equation}
\left| \Phi  \right\rangle  \equiv \left| {\overline \Psi  \left(
\bf y \right)\Psi \left( {\bf y}\prime \right)} \right\rangle  =
\overline \psi \left( \bf y \right)\exp \left(
{iq\int\limits_{{\bf y}\prime}^{\bf y} {dz^i } A_i \left( z
\right)} \right)\psi \left({\bf y}\prime \right)\left| 0
\right\rangle, \label{KR60}
\end{equation}
where $\left| 0 \right\rangle$ is the physical vacuum state and
the line integral appearing in the above expression is along a
spacelike path starting at ${\bf y}\prime$ and ending $\bf y$, on
a fixed time slice. It is worth noting here that the strings
between fermions have been introduced in order to have a
gauge-invariant function $\left| \Phi  \right\rangle $. In other
terms, each of these states represents a fermion-antifermion pair
surrounded by a cloud of gauge fields sufficient to maintain gauge
invariance. As we have already indicated, the fermions are taken
to be infinitely massive (static).

From our above discussion, we see that $\left\langle H
\right\rangle _\Phi$ reads
\begin{equation}
\left\langle H \right\rangle _\Phi   = \left\langle \Phi
\right|\int {d^3 x\left\{ { - \frac{1}{2}\Pi _i \left( {1 +
\frac{{e^2 }}{{\Delta ^2  + m^2 }}} \right)^{ - 1} \Pi ^i  +
\frac{1}{4}F_{ij} \left( {1 + \frac{{e^2 }}{{\Delta ^2  + m^2 }}}
\right)F^{ij} } \right\}}\left| \Phi  \right\rangle. \label{KR55}
\end{equation}
Consequently, we can write
Eq.(\ref{KR55}) as
\begin{equation}
\left\langle H \right\rangle _\Phi   = \left\langle \Phi
\right|\int {d^3 } \left\{ { - \frac{1}{2}\Pi _i \left( {1 -
\frac{{e^2 }}{{\nabla ^2  - m^2 }}} \right)^{ - 1} \Pi ^i }
\right\}\left| \Phi  \right\rangle, \label{KR65}
\end{equation}
where, in this static case, $\Delta ^2 = - \nabla ^2$. Observe that when $e=0$ we obtain the pure Maxwell theory, as mentioned after (\ref{KR15}).  From now on we will suppose $e\neq 0$.

Next, from our above Hamiltonian analysis, we note that
\begin{equation}
\Pi _i \left( x \right)\left| {\overline \Psi  \left( {\bf y}
\right)\Psi \left( {\bf y^\prime} \right)} \right\rangle  =
\overline \Psi \left( {\bf y} \right)\Psi \left( {\bf y^\prime}
\right)\Pi _i \left( x \right)\left| 0 \right\rangle  +
q\int\limits_{\bf y}^{\bf y^\prime} {dz_i \delta ^{(3)} \left(
{{\bf z} - {\bf x}} \right)\left| \Phi \right\rangle
}.\label{KR70}
\end{equation}
As a consequence, Eq.(\ref{KR65}) becomes
\begin{equation}
\left\langle H \right\rangle _\Phi   = \left\langle H
\right\rangle _0  + V^{\left( 1 \right)}  + V^{\left( 2 \right)},
\label{KR75}
\end{equation}
where $\left\langle H \right\rangle _0  = \left\langle 0
\right|H\left| 0 \right\rangle$. The $V^{\left( 1 \right)}$ and
$V^{\left( 2 \right)}$ terms are given by:
\begin{equation}
V^{\left( 1 \right)}  =  - \frac{{q^2 }}{2}\int {d^3 x} \int_{\bf
y}^{\bf y^\prime} {dz^\prime_i } \delta ^{\left( 3 \right)} \left(
{x - z^\prime} \right)\frac{1}{{\nabla _x^2  - M^2 }}\nabla _x^2
\int_{\bf y}^{\bf y^\prime} {dz^i } \delta ^{\left( 3 \right)}
\left( {x - z} \right), \label{KR80}
\end{equation}
and
\begin{equation}
V^{\left( 2 \right)}  =   \frac{{q^2 m^2}}{2}\int {d^3 x}
\int_{\bf y}^{\bf y^\prime} {dz^\prime_i } \delta ^{\left( 3
\right)} \left( {x - z^\prime} \right)\frac{1}{{\nabla _x^2  - M^2
}} \int_{\bf y}^{\bf y^\prime} {dz^i } \delta ^{\left( 3 \right)}
\left( {x - z} \right), \label{KR85}
\end{equation}
where $ M^2\equiv{m^2+ e^2} $ and the integrals over $z^i$ and
$z^\prime_i$ are zero except on the contour of integration.

The $V^{\left( 1 \right)}$ term may look peculiar, but it is
nothing but the familiar Yukawa interaction plus self-energy
terms. In effect, as was explained in Ref. \cite{GG2}, the
expression (\ref{KR80}) can also be written as
\begin{equation}
V^{\left( 1 \right)}  = \frac{{e^2 }}{2}\int_{\bf y}^{{\bf
y}^{\prime}  } {dz_i^{\prime}}\partial _i^{z^{\prime}} \int_{\bf
y}^{{\bf y}^{\prime}} {dz^i }\partial _z^i G\left( {{\bf
z}^{\prime},{\bf z}} \right), \label{KR90}
\end{equation}
where $G$ is the Green function
\begin{equation}
G({\bf z}^{\prime}  ,{\bf z}) = \frac{1}{{4\pi }}\frac{{e^{ -
M|{\bf z}^{\prime} - {\bf z}|} }}{{|{\bf z}^{\prime} - {\bf z}|}}.
\label{KR95}
\end{equation}
Employing Eq. (\ref{KR95}) and remembering that the integrals over
$z^i$ and $z_i^{\prime}$ are zero except on the contour of
integration, the expression (\ref{KR90}) reduces to the
Yukawa-type potential after subtracting the self-energy terms,
that is,
\begin{equation}
V^{\left( 1 \right)}  =  - \frac{{q^2 }}{{4\pi }}\frac{{e^{ -
M|{\bf y} - {\bf y}^ {\prime}| } }}{{|{\bf y} - {\bf
y}^{\prime}|}}. \label{KR100}
\end{equation}

We now turn our attention to the calculation of the $V^{\left( 2
\right)}$ term, which is given by
\begin{equation}
V^{\left( 2 \right)}  = \frac{{q^2 m^2 }}{2}\int_{\bf y}^{{\bf
y}^{\prime}  } {dz^{{\prime} i} } \int_{\bf y}^{{\bf y}^{\prime} }
{dz^i } G({\bf z}^{\prime} ,{\bf z}). \label{KR105}
\end{equation}
It is appropriate to observe here that the above term is similar
to the one found for the system consisting of a gauge field
interacting with a massive axion field \cite{GG2}.
Notwithstanding, in order to put our discussion into context it is
useful to summarize the relevant aspects of the calculation
described previously \cite{GG2}. In effect, as was explained in
Ref. \cite{GG2}, by using the Green function (\ref{KR95}) in
momentum space
\begin{equation}
\frac{1}{{4\pi }}\frac{{e^{ - M|{\bf z}^{\prime}   - {\bf z}|}
}}{{|{\bf z}^{\prime} - {\bf z}|}} = \int {\frac{{d^3 k}}{{\left(
{2\pi } \right)^3 }}\frac{{e^{i{\bf k} \cdot \left( {{\bf
z}^{\prime}- {\bf z}} \right)} }}{{{\bf k}^2  + M^2 }}},
\label{KR110}
\end{equation}
the expression (\ref{KR105}) can also be written as
\begin{equation}
V^{\left( 2 \right)}   = q^2 m^2 \int {\frac{{d^3 k}}{{\left(
{2\pi } \right)^3 }}} \left[ {1 - \cos \left( {{\bf k} \cdot {\bf
r}} \right)} \right]\frac{1}{{({\bf k}^2  + M^2)
}}\frac{1}{{\left( {{\bf {\hat n}} \cdot {\bf k}} \right)^2 }},
\label{KR115}
\end{equation}
where ${\bf {\hat n}} \equiv \frac{{{\bf y} - {\bf
y}^{\prime}}}{{|{\bf y} - {\bf y}^{\prime}| }}$ is a unit vector
and ${\bf r}={\bf y}-{\bf y^{\prime}}$ is the relative vector
between the quark and antiquark. Since ${\bf {\hat n}}$ and ${\bf
r}$ are parallel, we get accordingly
\begin{equation}
V^{\left( 2 \right)}   = \frac{{q^2 m^2 }}{{8\pi ^3
}}\int\limits_{ - \infty }^\infty {\frac{{dk_r }}{{k_r^2 }}}
\left[ {1 - \cos \left( {k_r r} \right)}
\right]\int\limits_0^\infty  {d^2 k_T \frac{1}{{(k_r^2 + k_T^2  +
M^2) }}}, \label{KR120}
\end{equation}
where $k_T$ denotes the momentum component perpendicular to ${\bf
r}$. We may further simplify Eq.(\ref{KR120}) by doing the $k_T$
integral, which leads immediately to the result
\begin{equation}
V^{\left( 2 \right)}  = \frac{{q^2 m^2 }}{{8\pi ^2 }}\int\limits_{
- \infty }^\infty  {\frac{{dk_r }}{{k_r^2 }}} \left[ {1 - \cos
\left( {k_r r} \right)} \right]\ln \left( {1 + \frac{{\Lambda ^2
}}{{k_r^2  + M^2 }}} \right), \label{KR125}
\end{equation}
where $\Lambda$ is an ultraviolet cutoff. We also observe at this
stage that similar integral was obtained independently in
Ref.\cite{Suganuma} in the context of the dual Ginzburg-Landau
theory by an entirely different approach.

We now proceed to compute the integral (\ref{KR125}). For this
purpose we introduce a new auxiliary parameter $\varepsilon$ by
making in the denominator of the integral (\ref{KR125}) the
substitution $k_r^2\rightarrow k_r^2+\varepsilon^2$. Thus it
follows that
\begin{equation}
V^{\left( 2 \right)} \equiv \lim _ {\varepsilon  \to 0}
{\widetilde V}^{\left( 2 \right)}= \lim _{\varepsilon \to
0}\frac{{q^2 m^2 }}{{8\pi ^2 }}\int\limits_{ - \infty }^\infty
{\frac{{dk_r }}{{(k_r^2  + \varepsilon ^2) }}} \left[ {1 - \cos
\left( {k_r r} \right)} \right]\ln \left( {1 + \frac{{\Lambda ^2
}}{{k_r^2  + M^2 }}} \right). \label{KR130}
\end{equation}
We further note that the integration on the $k_r$-complex plane
yields
\begin{equation}
{\widetilde V}^{\left( 2 \right)} = \frac{{q^2 m^2 }}{{8\pi
}}\left( {\frac{{1 - e^{ - \varepsilon |{\bf y} - {{\bf
y}^\prime}|  } }}{\varepsilon }} \right)\ln \left( {1 +
\frac{{\Lambda ^2 }}{{M^2 - \varepsilon ^2 }}} \right).
\label{KR135}
\end{equation}
Taking the limit $\varepsilon  \to 0$, expression (\ref{KR135})
then becomes
\begin{equation}
V^{\left( 2 \right)}  = \frac{{q^2 m^2 }}{{8\pi }}|{\bf y} - {{\bf
y}^\prime}| \ln \left( {1 + \frac{{\Lambda ^2 }}{{M^2 }}} \right).
\label{KR140}
\end{equation}
This, together with Eq.(\ref{KR100}), immediately shows
that the potential for two opposite charges located at ${\bf y}$
and ${\bf y^\prime}$ is given by
\begin{equation}
V(L) =  - \frac{{q^2 }}{{4\pi }}\frac{{e^{ - ML} }}{L} +
\frac{{q^2 m^2 }}{{8\pi }}L\ln \left( {1 + \frac{{\Lambda ^2
}}{{M^2 }}} \right), \label{KR145}
\end{equation}
where $L\equiv|{\bf y}-{\bf {y^\prime}}|$.
In this context it may be recalled the calculation reported in
Ref. \cite{Kondo} by taking into account topological nontrivial
sectors in $U(1)$ gauge theory is given by
\begin{equation}
V\left( L \right) =  - \frac{{q^2 }}{{4\pi }}\frac{1}{L} + \sigma
L. \label{KR155}
\end{equation}
Notice that the result (\ref{KR145}) agrees with (\ref{KR155}) in
the limit of large $M$. Thus one is led to the conclusion that,
although both calculations lead to confinement, the physical
mechanism of obtaining a linear potential is quite different. In
other terms, our result it may be considered as a physical
realization of the topological nontrivial sectors studied in
\cite{Kondo}. Let us also mention here that the result
(\ref{KR145}) is exactly the one obtained in Ref. \cite{Suganuma}
in the context of the dual Landau-Ginzburg theory. But we do not
think that the agreement is an accidental coincidence. As we
mentioned before, a gauge theory in the presence of external
fields and axions displays the same behavior \cite{GG2}. It seems
a challenging work to extend to a class of models the above
analysis, which can predict the same interaction energy. We expect
to report on progress along these lines soon.

\section{Final Remarks}

We have studied the confinement versus screening issue for a pair of antisymmetric tensors coupled to topological defects that eventually condense, giving a specific realization of the Julia--Toulouse phenomenon. We have seen that the Julia--Toulouse mechanism for a couple of massless antisymmetric tensors is responsible for the appearance of mass and the jump of rank in the magnetic sector while the electric sector becomes a BF--type coupling. The condensate absorbs and replaces one of the tensors and becomes the new massive propagating mode but does not couple directly to the probe charges.  The effects of the condensation are however felt through the BF coupling with the remaining massless tensor.  It is therefore not surprising that they become manifest in the interaction energy for the effective theory.  We have obtained the effective theory for the condensed phase in general and computed the interaction energy between two static probe charges, in a specific example, in order to test the confinement versus screening properties of the effective model. Our results show that the interaction energy in fact contains a linear confining term and an Yukawa type potential. It can be observed that confinement completely disappears in the limit $m\to 0$ while the screening takes over controlled by the topological mass parameter instead. Although we have considered the case where the effective model consists of the BF--coupling between a Kalb-Ramond field (that represents the condensate) and a Maxwell field, our results seem to be quite general.  A direct calculation for tensors of arbitrary rank in the present approach is however a quite challenging problem that we hope to be able to report in the future.

\section{ACKNOWLEDGMENTS}

One of us (CW) would like to thank the Physics Department of the
Universidad T\'{e}cnica F. Santa Mar\'{\i}a for the invitation and
hospitality during the earlier stages of this work.

\end{document}